\documentclass{article} 

\usepackage{verbatim}
\usepackage{amsthm}
\usepackage{url}

\usepackage{indentfirst}

\usepackage{hyperref} 

\usepackage{arxiv}

\usepackage[utf8]{inputenc} 
\usepackage[T1]{fontenc}    
\usepackage{booktabs}       
\usepackage{amsfonts}       
\usepackage{nicefrac}       
\usepackage{microtype}      
\usepackage{lipsum}
\usepackage{graphicx}

\newtheorem{thm}{Theorem}

\title{Publish, subscribe, and federate!}

\author{Marco Aur\'elio Spohn \\
Federal University of Fronteira Sul\\
Chapec\'o/SC, Brazil \\
\texttt{marco.spohn@uffs.edu.br}
}  

\begin{document}
 
\maketitle

\begin{abstract}
 Connecting usual things/objects to the Internet allows the monitoring and control of such things from anywhere, which is usually referred to as the Internet of Things (IoT). Things communicate among themselves or with other entities (\textit{e.g.}, a server) so that information can be gathered from things whilst proper actions can be taken upon them. A prominent communication approach adopted by many IoT applications is related to the Publish/Subscribe (P/S) paradigm. Any communication entity willing to provide some data announces its intention to a server (broker), establishing itself as a publisher for such data/topic. Entities that are willing to receive any published data, register themselves to the broker as subscribers. While employing just one broker might lead to a bottleneck and a single point of failure, when having multiple brokers one could end up having difficulties with their management. This work presents a scalable and efficient proposal for the federation of independent brokers, by allowing subscribers to get all their publications no matter to which broker publishers and subscribers are associated with.
\end{abstract}


\section{Introduction}

The capacity of connecting simple objects/things to the Internet allows the monitoring and control of such things. The result of a myriad of things connected to the Internet is called the Internet of Things (IoT)~\cite{iot}. Many applications comprising the IoT have adopted some sort of publish/subscribe (P/S) communication protocol. 
The P/S paradigm works as follows: i) an entity/thing willing to provide some information (called publication) signals that to a server/broker, becoming a publisher; ii) an entity (another thing or an external user/application) willing to receive a publication subscribes to it through the server/broker. 

As publications are sent to the broker, it sends them out to all active subscribers. When there is a single broker, we have a potential bottleneck and a single point of failure as major drawbacks. One prominent representative of the P/S paradigm with single broker is  the Message Queuing Telemetry Transport (MQTT)~\cite{mqtt} protocol, and it has been broadly adopted in IoT development platforms.

The problem in hand is to provide means for the federation of independent brokers. A federation is essentially an attempt to bring together similar parts so that they can perform a common task. As a federation, subscribers and publishers should be allowed to behave in a very similar way as there were just one broker; however, publications must now reach subscribers no matter to which broker they are associated with.  

Given that there are plenty of options for deploying an overlay Peer-to-Peer (P2P) network, one can assume that virtually interconnecting brokers can be taken for granted. The problem reduces then to the management of the federated brokers, as well as the routing of data packets amongst them.

In order to scale, the federation management process should incur a minimum overhead. Routing within the federated structure must be kept simple and efficient. For that reason, we resort to a multicast approach for dealing with the management and routing issues. 

To the best of our knowledge, this is the first work to present some foundations for the federation of autonomous brokers. Next, we present some of the works which target the drawbacks resulting from having a single broker. After that, we describe our proposal for the federation of brokers. Considering that the implementation of a full realization of the federation in an IoT platform is more complex, as well as composed of stages, we continue with an open issues section. Conclusions are then presented, including pointers for future works.

\section{Related works}\label{related-works}

Distributed Publish \& Subscribe for the Internet of Things (DPS)~\cite{intel-dps} is a distributed implementation for the P/S paradigm. DPS builds a mesh network mainly around subscribers, accomplishing routing from publishers to subscribers based on subscription topics, and allowing IP multicast in local subnets. The main difference between DPS and our solution is that DPS does not provide means for the federation of autonomous brokers, but instead it focus on the mesh deployment centered on publishers and subscribers, without multiple independent brokers. 

Direct Multicast-MQTT (DM-MQTT)~\cite{dm-mqtt} provides improvements to MQTT to mitigate the problems resulting from having just a single centralized broker. As we increase the number of subscribers, the number of messages sent out by the broker increases proportionally, resulting in higher delays. To reduce data transfer delays, DM-MQTT resorts to Software Defined Network (SDN) multicast trees connecting publishers to subscribers, bypassing the centralized broker. Once more, compared to our proposal, their solution is not a federation of autonomous brokers.

\section{Federation}\label{federation}

When working to solve a problem, one knows that simplicity is desired from the very beginning. The problem we are looking at concerns the federation of independent P/S brokers interconnected through an overlay/virtual network. Making it simple implies on adopting a reduced, but efficient, set of  control mechanisms for bringing and maintaining together a group of P/S brokers willing to assemble as a federation. The main result from such federation is that brokers should be able to handle any content from publishers to subscribers no matter to which broker they are associated with. 

As a side effect, the federation itself adds an extra overlay layer which can provide means to explore redundancy, making it possible to resort to lighter transport protocols (\textit{e.g.}, UDP) when conveying control and data messages. A mesh topology would come well to address such requirements, assuming that the underlying virtual topology provides the required properties (\textit{i.e.}, multiple paths between pairs of nodes).

Control communication among federated brokers must be devised for scaling in terms of an increasing number of endpoint entities (\textit{i.e.}, publishers and subscribers) and brokers. Publishers and subscribers should behave as usual, as there were just the broker they are associated with. 

For our proposal, we borrow the fundamental communication principles from a particular multicast protocol designed for Mobile Ad Hoc Networks (MANETs)~\cite{manets}. In such networks, it is known there is no fixed communication infrastructure, leaving to the \textit{ad hoc} members themselves the task of building and keeping an autonomous communication infrastructure. In such an adversarial environment, it is paramount that communication protocols incur less control overhead when compared to their counterparts from traditional communication infrastructures. To make things worse, one can not count on an accurate view of the network topology due to node mobility, usually unpredictable.

We borrow from the \textit{Protocol for Unified Multicasting through Announcements} (PUMA)~\cite{puma} the idea surrounding the simple mechanism for establishing a mesh structure around multicast receivers: the multicast announcements.
In our solution, besides brokers playing  a similar role as receivers in PUMA, brokers have a direct connection to
senders (publishers), while it is not an intrinsic assumption in PUMA where the multicast mesh is built exclusively around receivers.

In PUMA, multicast meshes are built around receivers, where an elected receiver (the one with the largest/smallest ID) acts as a core/leader for building and maintaining the mesh structure. The core starts by broadcasting an announcement to the whole network, not targeting only the receivers. For a MANET, flooding the network is a straightforward way to likely reach all nodes, assuming that the network topology is not known in advance, much less who and where exactly the receivers are. Considering that multicast announcements are sent 
out periodically by the core, they eventually reach all receivers which, on their turn, learn the current shortest distance to the core. Depending on the predefined mesh redundancy degree, mesh members (\textit{i.e.}, receivers, and intermediary nodes connecting receivers towards the core) set as their parents in the mesh as many neighbors as the defined redundancy. The control overhead comes down to just one message (\textit{i.e.}, the multicast announcement) started out by the core periodically; therefore, it is quite scalable.

PUMA was designed for a scenario with node mobility, no fixed communication infrastructure, and wireless omnidirectional communication: when a node transmits, all of its neighbors can potentially receive the transmission. In our case, brokers are assumed to be deployed in an infrastructure-based network, such as the Internet. Therefore, the fact that a single transmission of a multicast announcement in PUMA can reach several nodes (neighbors) at the same time is not at stake in our particular environment/scenario.

\subsection{Mesh deployment}

We evolve the multicast announcement concept by defining a relationship between multicast groups and topics of interest. It is known that publications to any given topic must reach all its subscribers, which on their turn should be allowed to be associated with any federated broker. The idea is to have one broker acting as the core for any set of related topics, building around itself the mesh required to the routing of the corresponding publications to all their subscribers. This way, there is a reduced control overhead for building and maintaining a routing infrastructure among brokers.

The following assumptions are taken for the creation and maintenance of the federation of brokers:
\begin{itemize}
 \item Brokers are connected over an overlay/virtual network previously established;
 \item Publishers and subscribers have knowledge of at least one reference broker;
 \item There is a mapping between a set of topics (defined by an application) and a multicast group;
 \item There is an out of band (management) multicast group spanning all the federated brokers. The corresponding mesh is deployed and maintained following the same procedure as any other regular application mesh.
\end{itemize}

Once a broker receives for the very first time a subscription request for a given topic, the broker checks if any other federated broker has already announced itself as the core for such a topic. The case when more than one broker assumes the core role is easily sorted out after a tie-breaker criterion is defined. The simplest criteria would be to elect as the core the broker with the largest (or smallest) identifier (ID). 

For the creation and maintenance of the virtual network topology, one could easily adopt or adapt, some of the many solutions employed for peer-to-peer (P2P) networks~\cite{overlay3}~\cite{overlay2}~\cite{overlay1}. The mappings between topics and multicast groups must be previously agreed upon. To support such mapping, there is a management multicast group which can be employed for advertising among brokers any mappings, even allowing the announcement of several mappings in a single advertisement. Applications can interact with brokers to request such mapping following some standard procedure, which must be well defined previously.

Whilst all brokers participate in the management group, the same is not necessarily true for the topic/application 
groups. Overall, a broker takes part of any topic group in the following cases: i) it has at least one subscriber for the related topic; or ii) the broker is an interconnecting node (\textit{i.e.}, the node interconnects at least one of its neighbors to the corresponding core). Control information is kept to a minimum, given that every broker should just register the multicast groups it belongs to. The routing decision is quite straightforward: i) if the broker is just an interconnecting node for the destination multicast address, the broker simply forwards the packet to its neighbors in the corresponding mesh (it learns that from the periodical multicast announcements); or, ii) besides forwarding the packet as previously, it sends out the packet to all the local subscribers. This way, publications can reach all the subscribers no matter to which broker they have been associated with.

A core announcement packet carries the following information: 
\begin{itemize}
\item Core ID; 
\item Distance to the core (initially zero, and it gets incremented before the packet is propagated away by a receiving neighbor); 
\item A mesh membership flag, with value $1$ when the sending node is a mesh member, and $0$ otherwise; 
\item and, alternatively, a list of parents (\textit{i.e.}, neighbor(s) leading to the core), which depends on the mesh redundancy degree. 
\end{itemize}

A broker is a mesh member when it has at least one local subscriber or it interconnects other mesh neighbors to the core. There are two possible ways to learn mesh membership when the broker is just an interconnecting node: the broker is listed as one of the neighbor's parents; or, the broker is at a shorter distance to the core compared to a mesh neighbor.

Besides the core announcement control packet, we introduce the mesh membership announcement tailored for notifying interconnecting neighbors not yet mesh members. When a core announcement comes from a neighbor which results in being chosen as a mesh parent for the current broker, it has to notify the neighbor so that it can take the proper actions. That is, a mesh membership announcement is sent out to the target neighbor (mesh parent) so that it sets itself as a mesh member as well. Mesh membership announcements can cascade up towards to the core, and they carry the core ID and the sequence number of the corresponding core announcement (alternatively, they can include the list of broker's parents).

In order to build and maintain the management group, its core can be predefined or elected (\textit{e.g.}, by choosing the broker with the smallest/largest ID as done in regular groups). The core starts by advertising itself as the management group leader by flooding the virtual network with a multicast announcement (periodically new announcements are started out with an increasing sequence number).

A core announcement (management or regular group core) providing a shorter or equal distance to the core (\textit{i.e.}, $<$announced distance + 1$>$ is equal to or smaller than current known distance) is processed as follows:

\begin{itemize}
    \item When necessary, updates the distance locally and, in case the mesh redundancy allows a new link/parent, sets the  sending neighbor as one of the broker's parents; 
   \item If the receiving broker is a mesh member, but the sending neighbor is not, and a mesh membership announcement was not yet sent to the neighbor for this particular core announcement (known by its sequence number), then a mesh membership announcement must be sent to the advertising neighbor; 
   \item In case of shortening the distance to the core, the broker sends out an updated core announcement (with current distance to the core, and proper mesh flag setting) to the remaining neighbors (\textit{i.e.}, all neighbors except the one from which the announcement was heard from).
\end{itemize}

A mesh membership announcement is processed by the receiving broker as follows:

\begin{itemize}
 \item If the broker is not yet a mesh member for the given core, the broker flags itself as a mesh member, and send a mesh membership announcement to all the broker's parents (preserving core ID and sequence numbering in the announcement). 
 \item The core itself is an exception, and it just registers the corresponding mesh neighbors (needed for the proper data routing process).
\end{itemize}

\begin{figure}
 \centering
 \includegraphics[width=0.7\columnwidth]{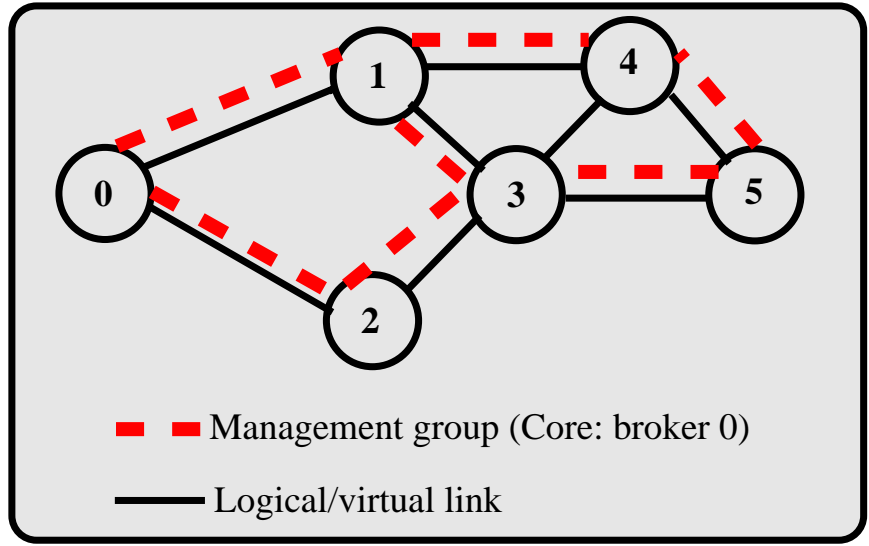}
 \caption{An example of a virtual topology with the management group (which includes all brokers).}
 \label{fig:management_group}
\end{figure}

It is important to understand that every broker learns about any existing core (\textit{i.e.}, management, and regular groups); however, a broker takes part in a particular mesh if it has to. Figure~\ref{fig:management_group} depicts an example considering a virtual network with six brokers, having node 0 as the core for the management group. In this example mesh redundancy is set to two, meaning that every broker has up to two neighbors leading to the core: the link between broker 3 and 4 is not in the mesh, because it does not provide a shortest distance to the core (in terms of number of hops).

\begin{figure*}
 \centering
 \includegraphics[width=0.7\textwidth]{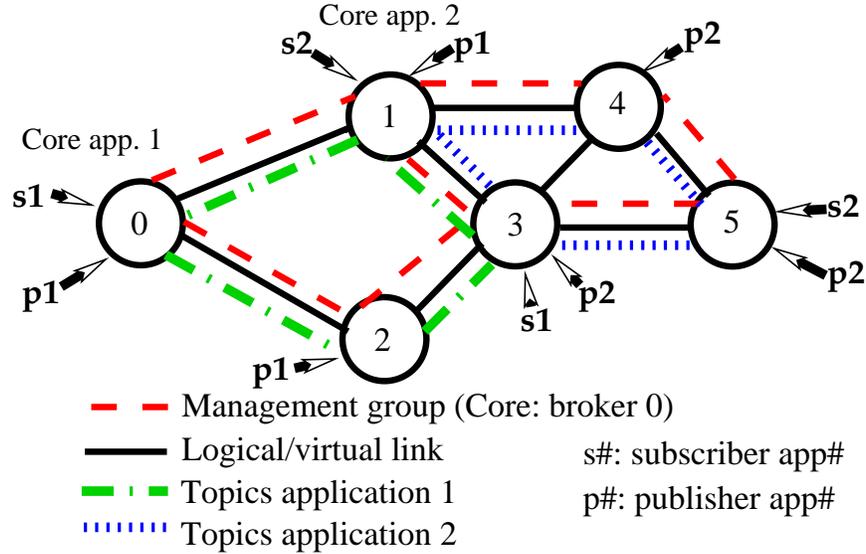}
 \caption{Example with two topic/application groups.}
 \label{fig:example_two_applications}
\end{figure*}

Figure~\ref{fig:example_two_applications} depicts a situation with two topics/applications groups: application 1 having as its core broker 0, and application 2 having as its core broker 1. We could assume that broker 0 received the first subscriber for a topic from application 1, while broker 1 had the first subscriber for a topic belonging to application 2. In a latter moment, broker 3 receives a subscription for application 1, and broker 5 receives a subscription for application 2.


\begin{figure*}
 \centering
 \includegraphics[width=0.8\textwidth]{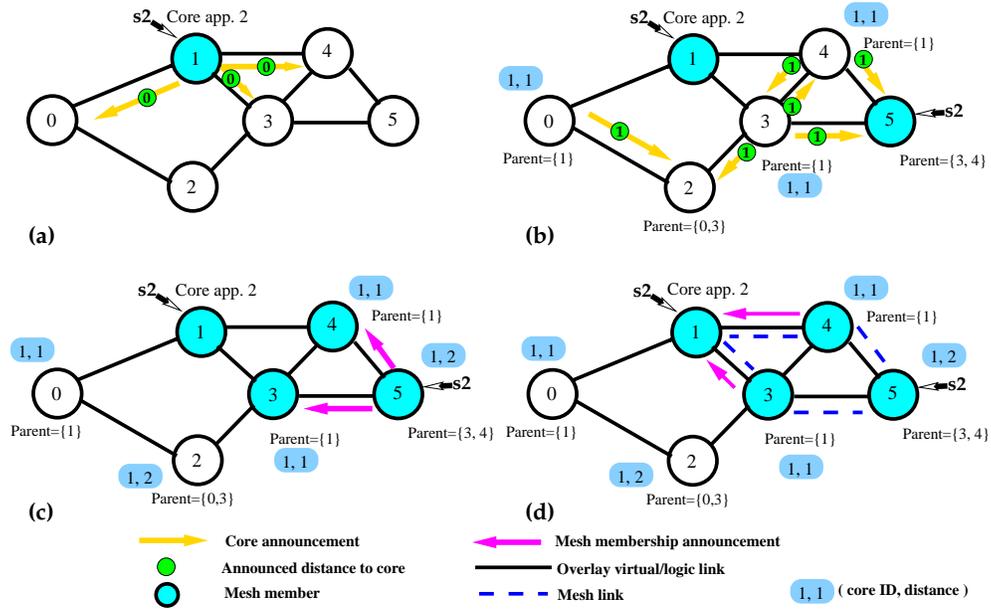}
 \caption{Mesh building process (redundancy = 2).}
 \label{fig:building_the_mesh}
\end{figure*}

Figure~\ref{fig:building_the_mesh} shows the process for setting up the mesh for application 2 depicted in Figure~\ref{fig:example_two_applications}. Assume that broker 1 is the first to receive a subscription for any topic belonging to application 2. It starts by announcing itself as the core for topics related to application 2. As the announcement traverses the network, brokers set their list of parents appropriately (assuming mesh redundancy equal to two). Brokers with no local subscribers just forward the core announcement to their neighbors. Broker 5 has one subscriber for the same application, and once receiving the core announcement it triggers a mesh membership announcement to be sent to each one of its parents (\textit{i.e.}, brokers 3 and 4). Upon receiving such announcement, brokers 3 and 4 set themselves as mesh members, because they are interconnecting brokers, and then send their mesh membership to their parents (in this case, the core itself).

Even though there may have more than one core for the same topics/application temporarily, the multicast announcement is sent out periodically, allowing to converge to the mesh starting from the core with the largest/smallest ID. The announcement periodicity must be properly addressed in the implementation of the federation in a real IoT platform.

\subsection{Routing}

\begin{figure*}
 \centering
 \includegraphics[width=0.8\textwidth]{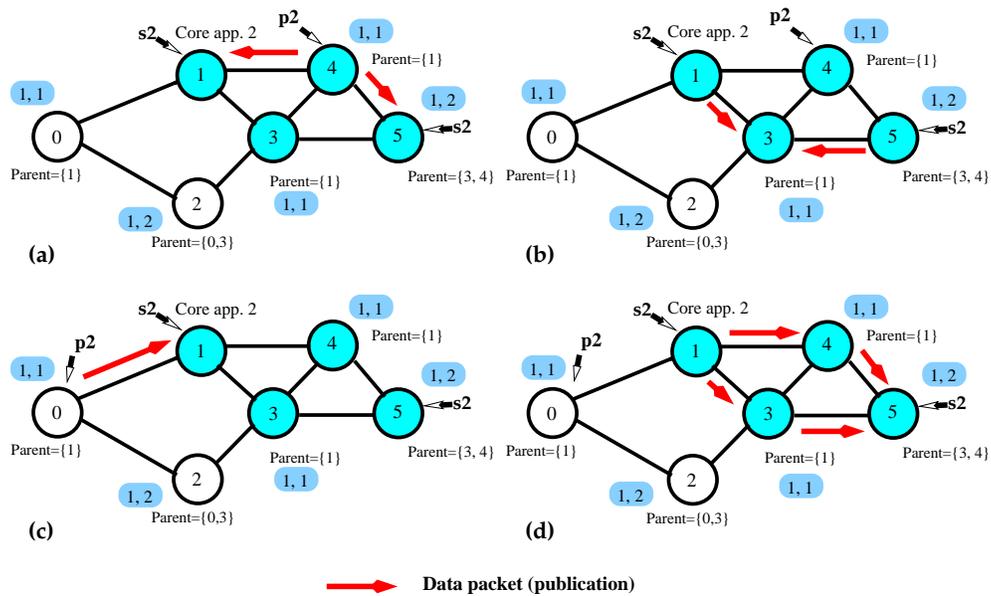}
 \caption{Data packet (topic publication) routing process.}
 \label{fig:routing}
\end{figure*}

Routing of publications are quite straightforward due to the mesh structure, having two possible situations: 
\begin{itemize}
\item If a publication starts at a mesh member, the packet is just forwarded to all mesh neighbors, after which it spreads along with the mesh (looping is avoided by checking packet ID). An example is shown in Figures~\ref{fig:routing}~(a)--(b), having a publisher associated with broker 4.
\item If the publisher is an outsider to the mesh, it is worth remembering that all brokers learn about existing cores and their active applications/groups; therefore, it just requires sending any publication towards the corresponding core, from which it spreads along with the mesh. An example is shown in Figures~\ref{fig:routing}~(c)--(d), having a publisher associated with broker 0, which is not a mesh member of application 2.
\end{itemize}

It is worth mentioning that, when a publication is sent from somewhere outside the mesh,  the publication does not necessarily need to go all the way to the core, as long as the packet reaches any mesh member while being forwarded towards the core.

\subsection{Formal properties}

\begin{thm}\label{theo:overhead}
For a network topology with \textbf{n} brokers and \textbf{l} links, core announcement overhead is bounded to \(O(l)\) messages.
\begin{proof}
A core announcement can start at any particular broker in the overlay network. An announcement traverses the same link at most twice when links connect pairs of brokers at the same distance from the originating core. Therefore, the total transmissions of an announcement is bounded to $2\times~l$, which is of order \(O(l)\). 
\end{proof}
\end{thm}

\begin{thm}\label{theo:liveness}
 For a network topology with n brokers and l links, the mesh structure converges after a finite period of time.
\begin{proof}
 This proof concerns the safety and liveness of the protocol; that is, it converges with no possibility of deadlocks. Given that Theorem~\ref{theo:overhead} proves that the core announcement overhead is bounded by \(O(l)\) transmissions and that each transmission takes a limited time, all brokers, including the intended receivers (\textit{i.e.}, brokers with subscribers), get to know the shortest distance to the core and the corresponding neighbors leading to the core (\textit{i.e.}, parent nodes). Brokers with subscribers notify all parents (bounded to the mesh redundancy), which are at the shortest distance to the core, through a mesh membership message. Interconnecting brokers become mesh members, allowing that all the required mesh membership announcements cascade back towards to the core, ending the mesh deployment process. Given that all mesh membership announcements are related to the corresponding core announcement (with a unique identifier), and that a mesh membership announcement is transmitted at most once by any mesh member, there is no possibility of deadlock.
\end{proof}
\end{thm}

\section{Open issues}\label{open-issues}

The main objective of this work is to present some foundations and basic algorithms for the federation of autonomous brokers, serving as a framework for the full realization of the federation in some specific P/S implementation. As so, many open issues arise, such as the following ones:

\begin{itemize}
 \item  Guidelines should be defined for choosing the desired topology of the overlay network, knowing that it shapes the resulting meshes and likely affecting performance metrics.
 \item One must properly address the mappings between sets of topics and multicast groups.
 \item Other management issues will arise when dealing with the specificities of any application. Such problems can be sorted out with the support from the management mesh, which encompasses all brokers.
 \item Even though control overhead is bounded and it is expected to be scalable, performance issues must be investigated in a large set of scenarios and applications. 
\end{itemize}

While such open issues can well be addressed in future works, it might be necessary to adjust the underlying algorithms appropriately. However, it is not an exclusive matter for the current proposal, but it equally concerns any evolving protocol.

\section{Conclusions}\label{conclusions}

The P/S communication paradigm is simple and well fitted to many IoT applications. When the P/S operations are provided by a single broker (\textit{e.g.}, as in MQTT), a bottleneck might surface, incurring unacceptable delays. Besides that, a single point of failure is an issue when some fault tolerance is required. On the other hand, by having multiple brokers one could somehow address such drawbacks. However, managing multiple brokers is not trivial.

In this work, we have proposed an approach for the federation of autonomous P/S brokers. From start, multicast routing among federated brokers was taken as one of the principles for designing a scalable solution. An efficient and low-cost multicast routing protocol from MANETs  (\textit{i.e.}, PUMA) was the basic inspiration for our solution. PUMA resorts to a single control packet (multicast announcement) for setting up and maintaining a mesh structure connecting all the intended receivers in a multicast group.

We assume that brokers are interconnected through an overlay network over infrastructure such as the Internet, instead of the regular scenario PUMA was designed for (\textit{i.e.}, node mobility,  no fixed communication infrastructure, and wireless omnidirectional transmissions). Nevertheless, to some extent, we were able to design a solution inheriting the principles of low control overhead and the mesh infrastructure present in PUMA, which in our case is built around the intended subscribers.

The concepts and basic algorithms for the federation of brokers were laid on in this work. Nevertheless, there is a need for a full realization of the proposal through a real or simulated implementation, as well as other open issues as illustrated previously. One promising future work relates to the federation of MQTT brokers, aiming at devising a usable solution with the minimum set of required modifications.

\bibliographystyle{unsrt} 
\bibliography{references}

\end{document}